\documentclass{article}
\usepackage[utf8]{inputenc}
\usepackage[T1]{fontenc}
\usepackage[english]{babel}
\usepackage{amsmath}
\usepackage{amssymb}
\usepackage{amsthm}
\usepackage{enumerate}
\usepackage{appendix}
\usepackage{geometry}
\usepackage{eurosym}
\usepackage{float}
\usepackage{hyperref}
\hypersetup{
    colorlinks,
    citecolor=blue,
    filecolor=blue,
    linkcolor=blue,
    urlcolor=blue
}
\usepackage{todonotes}
\usepackage{mathrsfs}
 \usepackage[round]{natbib}

 \usepackage{rotating}

\def\be{\begin{eqnarray}}
\def\ee{\end{eqnarray}}
\def\b*{\begin{eqnarray*}}
\def\e*{\end{eqnarray*}}

\newtheorem{Proposition}{Proposition}[part]

\newcommand{\ba}{\begin{array}}
\newcommand{\ea}{\end{array}}
\newcommand{\ben}{\begin{equation*}} 
\newcommand{\een}{\end{equation*}}
\newcommand{\bea}{\begin{eqnarray}}
 \newcommand{\eea}{\end{eqnarray}}
\newcommand{\bean}{\begin{eqnarray*}} 
\newcommand{\eean}{\end{eqnarray*}}
\newcommand{\bel}{\begin{align}} 
\newcommand{\eel}{\end{align}}
\newcommand{\beln}{\begin{align*}} 
\newcommand{\eeln}{\end{align*}}
\newcommand{\bit}{\begin{itemize}}
\newcommand{\eit}{\end{itemize}}

\makeatletter \@addtoreset{equation}{section}

\@addtoreset{Definition}{section}

\@addtoreset{Theorem}{section}

\@addtoreset{Proposition}{section}

\@addtoreset{Property}{section}

\@addtoreset{Assumption}{section}

\@addtoreset{Corollary}{section}

\@addtoreset{Lemma}{section}

\@addtoreset{Remark}{section}

\@addtoreset{Example}{section}

\@addtoreset{Condition}{section}

\def \E{\mathbb{E}}

\def \H{\mathbb{H}}
\def \L{\mathbb{L}}
\def \M{\mathbb{M}}
\def \N{\mathbb{N}}
\def \P{\mathbb{P}}
\def \Q{\mathbb{Q}}
\def \R{\mathbb{R}}

\def \Z{\mathbb{Z}}

\def \G{\mathbb{G}}

\def\={\;=\;}
\def\.{\;.}

\def \i{1\!\mbox{\rm I}}
\def\1{{\bf 1}}

 \def\normeL2#1{\left\|{#1}\right\|_{L^2}}
 
\newcommand{\alias}[2]{
\providecommand{#1}{}
\renewcommand{#1}{#2}
}

\alias{\P}{\mathbb{P}}
\alias{\N}{\mathcal{N}}
\alias{\L}{\mathcal{L}}
\alias{\Z}{\mathbb{Z}}
\alias{\Q}{\mathbb{Q}}
\alias{\R}{\mathbb{R}}
\alias{\C}{\mathcal{C}}
\alias{\T}{\mathbb{T}}
\alias{\E}{\mathbb{E}}
\alias{\H}{\mathcal{H}}
\alias{\1}{\mathbf{1}}
\alias{\B}{\mathcal{B}}
\alias{\M}{\mathcal{M}}
\alias{\G}{\mathcal{G}}
\alias{\Y}{Y_{\bullet}}

\newcommand{\nc}{\newcommand}
\nc{\cA}{{\mathcal A}} \nc{\cB}{{\mathcal B}} \nc{\cC}{{\mathcal
C}} \nc{\cD}{{\mathcal D}} \nc{\bbD}{\mathbb{D}}
\nc{\cG}{{\mathcal G}} \nc{\cF}{{\mathcal F}} \nc{\cS}{{\mathcal
S}} \nc{\cU}{{\mathcal U}} \nc{\cH}{{\mathcal H}}
\nc{\cK}{{\mathcal K}} \nc{\cM}{{\mathcal M}} \nc{\cO}{{\mathcal
O}} \nc{\cP}{{\mathcal P}} \nc{\bbE}{\mathbb{E}}
\nc{\bbEP}{\mathbb{E}_{\mathbb{P}}}\nc{\bbL}{\mathbb{L}}
\nc{\bbP}{\mathbb{P}} \nc{\bbQ}{\mathbb{Q}} \nc{\del}{\partial}
\nc{\Om}{\Omega} \nc{\om}{\omega} \nc{\bbR}{\mathbb{R}}
\nc{\bbC}{\mathbb{C}} \nc{\bfr}{\begin{flushright}}
\nc{\efr}{\end{flushright}} \nc{\dXt}{\delta q_{t}}
\nc{\dXs}{\delta q_{s}} \nc{\bs}{\blacksquare} \nc{\dX}{\delta q}
\nc{\dY}{\Delta Y}
\nc{\dnkx}{\left(X(T^{n}_{k})-X(T^{n}_{k-1})\right)}
\nc{\esssup}{\mathrm{ess}\mbox{ }\mathrm{sup}}
\nc{\essinf}{\mathrm{ess}\mbox{ } \mathrm{inf}}
\nc{\dhats}{\widehat{\delta_s}}

\nc{\chf}{\mbox{$\mathbf1$}}
\nc{\ind}{\mathds{1}}

\begingroup\expandafter\expandafter\expandafter\endgroup
\expandafter\ifx\csname pdfsuppresswarningpagegroup\endcsname\relax
\else
\fi
\graphicspath{{./fig/}}

\nc{\mum}{ \mu_{\rm m} }
\nc{\muv}{ \mu_{\rm v} }
\nc{\mumv}{ \mu_{\rm mv} }
\nc{\Hm}{ H_{\rm m} }
\nc{\Hv}{ H_{\rm v} }
\nc{\dd}{ {\rm d} }

\begin{document}
\title{Optimal Incentive for Regulated Production}
\author{Benhao Du\thanks{University Paris Dauphine --- PSL Research University.}\quad Thomas Treillard\thanks{University Paris Dauphine --- PSL Research University.}\quad Francois Wang\thanks{University Paris Dauphine --- PSL Research University.}}
\maketitle

\begin{abstract}
This paper explores stochastic control models in the context of decarbonization within the energy market. We study three progressively complex scenarios: (1) a single firm operating with two technologies—one polluting and one clean,(2)two firms model and (3) two  firms without any regulatory incentive. For each setting, we formulate the corresponding stochastic control problem and characterize the firms' optimal strategies in terms of investment and production. The analysis highlights the strategic interactions between firms and the role of incentives in accelerating the transition to cleaner technologies.
\end{abstract}

\noindent{\bf keywords}: Stochastic control in the energy market,HJB, PDE.

\tableofcontents

\section{Introduction}
The transition toward a low-carbon economy is a central challenge in modern energy policy. In the energy sector, firms often face trade-offs between profitability and environmental sustainability, particularly when choosing between polluting and clean technologies. Understanding how firms make such decisions under uncertainty and competition is crucial for designing effective regulatory frameworks.

Stochastic control provides a natural mathematical framework to model dynamic decision-making under uncertainty. In this paper, we examine three stylized models of decarbonization using tools from stochastic control theory:

First, we analyze a single firm that can choose between two technologies—a traditional, polluting one and a cleaner but potentially costlier alternative. The firm optimizes its production and technology-switching policy under uncertain demand or price conditions.

Second, we introduce a centralized incentive mechanism (e.g., a subsidy or carbon tax) and study how it alters the firms’ optimal strategies and the resulting equilibrium.

Third, we extend the model to two competing firms where each firm's decisions influence market outcomes. In this version, no external incentives are provided, leading to inefficient levels of decarbonization due to strategic behavior.

Our goal is to compare these three frameworks and assess the conditions under which decarbonization emerges as a rational and optimal outcome. We aim to quantify the impact of both competition and regulation on investment decisions in clean technologies.






\section{Model}

\subsection{Single firm model}
In this section, we analyze the model of decarbonation in the situation where a single firm would control the 2 technologies $X^1$ and $X^2$, the dynamic of each technology $i=1,2$ is given by
\begin{align*}
    &dX_t^i=a_t^idt+\sigma_idW_t^i
\end{align*}
with two independent Brownian motion $W^1,W^2$. 
The cost for Firm is given by
\begin{align*}
    &c(a)=\frac{1}{2}\frac{a_1^2}{\gamma_1}+\frac{1}{2}\frac{a_2^2}{\gamma_2}
\end{align*}
The objective for the firm with is to maximize its expected utility by choosing its control strategy $a=(a^1,a^2)$
\begin{align*}
    &J_A=E[U_A(Y_T-\int_0^T (f(X_t)-c(a_t))dt)]\\
    &f(x)=(p_0-p_1x_1+p_2x_2)(x_1+x_2)
\end{align*}
The objective for the principal is to maximize its expected utility .
\begin{align*}
    &J_p=E[U_p(-Y_T-\int_0^T g(X_t)dt)]\\
    &dY_t=Z^1_tdX^1_t+Z^2_tdX^2_t+\frac{1}{2}\eta_A(\sigma_1^2(Z^1_t)^2+\sigma_2^2(Z^2_t)^2)dt-H_A(Z_t,X_t)dt\\
    &H_A(Z,X)=\hat{a}_1(Z_1)(Z_1)+\hat{a}_2(Z_2)Z_2-c(\hat{a}(Z))+f(X)\\
    &g(X)=\frac{1}{2}\kappa (X^1)^2+\lambda (X^1+X^2-\delta)^2\\
    &\hat{a}_1(z_1)=\gamma_1 z_1,\hat{a}_2(z_2)=\gamma_2 z_2
\end{align*}
We assume that there is no interaction between $a_1$ and $a_2$, so the Principal's optimal control problem becomes:
\begin{align*}
    &V(0,x,y)=\sup_Z E[U_p(-Y_T -\int_0^Tg(X_t)dt)]\\
    &dY_t=\sigma^1dW_t^1+\sigma^2dW_t^2+\frac{1}{2}\eta_A(\sigma_1^2(Z^1_t)^2+\sigma_2^2(Z^2_t)^2)dt+c(\hat{a}(Z_t))dt-f(X_t)dt\\
\end{align*}
\subsection{Two firms model}

\begin{align*}
&\dd X^i_t = a^i_t \dd t + \sigma_i \dd W^i_t, \quad 
J_i := \E\Big[ U_i \Big( Y^i_T  + \int_0^T f_i(X_t) - c_i(a^i_t) \big) \dd t \Big)\Big], \quad
f_i(x) := (p_0 - p_i x_i - p_j x_j), \quad c_i(a) := \frac1{2\gamma_i} a^2, \\
& J_{\rm P} := \E\Big[ U_i \Big( - Y_T  - \int_0^T g(X_t) \big) \dd t \Big)\Big], \quad
Y_T := Y^1_T + Y^2_T, \quad
g(x) := \frac12 \kappa x_2^2 + \frac12 \lambda (x_1+x_2-\delta)^2, \\
& \dd Y^i_t = Z^{i,i}_t \dd X^i_t + Z^{i,j}_t \dd X^j_t + \frac12 \eta_i\big[ \sigma_i^2 (Z^{i,i}_t)^2 + \sigma_j^2 (Z^{i,j}_t)^2\big] \dd t
- H_i(Z_t,X_t) \dd t, \\
& H_i(z,x) := \hat a_i(z_{ii}) z_{ii} - c_i(\hat a_i(z_{ii})) + \hat a_j(z_{jj}) z_{ij} + f_i(x)
\end{align*}
From which we can write
\begin{align*}
& \dd Y^i_t =  \big[c_i(\hat a_i(Z^{ii}_t)) - f_i(X_t) + \frac12 \eta_i\big( \sigma_i^2 (Z^{ii}_t)^2 + \sigma_j^2 (Z^{i,j}_t)^2\big)\big]
+ \sigma_i Z^{ii}_t  \dd W^i_t + \sigma_j Z^{i,j}_t \dd W^j_t
\end{align*}
And thus for the sum of the payments $Y$ we get
\begin{align*}
& \dd Y_t =  \big[c(\hat a(Z_t)) - f(X_t) +  R(Z_t) \big)\big] + \hat \sigma_1(Z_t)  \dd W^1_t + \hat \sigma_2(Z_t) \dd W^2_t, \\
& c(\hat a(z)) := c_1(\hat a_1(z_{11})) +  c_2(\hat a_2(z_{22})) = \frac12\big( \gamma_1 z_{11}^2 + \gamma_2 z_{22}^2\big), \\
& f(x) := f_1(x) + f_2(x) = (p_0 - p_1 x_1 - p_2 x_2) (x_1 + x_2), \\
& R(z) := \frac12 \big( \eta_1 \sigma_1^2 z_{11}^2 +\eta_1 \sigma_2^2 z_{12}^2 + \eta_2 \sigma_2^2 z_{22}^2 + \eta_2 \sigma_1^2 z_{21}^2\big), \\
& \hat \sigma_1(z) := (z_{11} + z_{21})\sigma_1, \quad 
\hat \sigma_2(z) := (z_{22} + z_{12})\sigma_2
\end{align*}
Hence, the problem of the regulator reads
\begin{align*}
&V(0,x,y) := \sup_{z} \E\Big[ U_{\rm P}\Big(-Y_T - \int_0^T g(X_t) \dd t\Big) \Big], \\
& \dd X^i_t = \hat a_i(Z^{ii}_t) \dd t + \sigma_i \dd W^i_t, \quad \hat a_i(z) := \gamma_i z_{ii}, \\
& \dd Y_t =  \big[c(\hat a(Z_t)) - f(X_t) +  R(Z_t) \big)\big] + \hat \sigma_1(Z_t)  \dd W^1_t + \hat \sigma_2(Z_t) \dd W^2_t, \\
\end{align*}

\subsection{Firms equilibrium without incentives}
In this section, we analyze the strategic interaction between two competing firms prior to any external intervention (e.g., by a regulator or principal). The goal is to determine the Nash equilibrium of the differential game modeling their behavior. We assume no external incentives are present, meaning each firm solely seeks to maximize its own expected utility based on its costs.\\

\noindent
Components of our model:
\begin{itemize}
    \item Players: Two firms, Firm 1 and Firm 2.
    \item State Variables:
        \begin{itemize}
            \item $X_t$: Represents the state of Firm 1 carbon emission rate at time $t$.
            \item $Y_t$: Represents the state of Firm 2 carbon emission rate at time $t$.
        \end{itemize}
    \item Strategies:
        \begin{itemize}
            \item $a^1_t$: Firm 1's abatement effort at time $t$.
            \item $a^2_t$: Firm 2's abatement effort at time $t$.
        \end{itemize}
    \item Dynamics:
        \begin{align*}
            \dd X_t &= a^1_t \, dt + \sigma_1 \, dW^1_t, \quad X_0 = x_0, \\
            \dd Y_t &= a^2_t \, dt + \sigma_2 \, dW^2_t, \quad Y_0 = y_0.
        \end{align*}
        $\sigma_i > 0$ are constant, $W^1_t, W^2_t$ independant Brownian motions.
        
    \item Cost of Action: The cost for Firm $i$ to exert effort $a^i_t$ is given by :
        \begin{itemize}
            \item Firm 1's cost: $c_1(a^1_t) = \frac{1}{2\gamma_1} (a^1_t)^2$
            \item Firm 2's cost: $c_2(a^2_t) = \frac{1}{2\gamma_2} (a^2_t)^2$
        \end{itemize}
        where $\gamma_1 > \gamma_2 > 0$.
        
    \item Payoff: The instantaneous net payoff for each firm at time $t$ depends on both firms' emission rates and the firm's own action cost. 
    \begin{itemize}
        \item Firm 1: $\pi_1(X_t, Y_t, a^1_t) = (p_0 - p_1 X_t - p_2 Y_t) X_t - \frac{1}{2\gamma_1} (a^1_t)^2$ 
        \item Firm 2: $\pi_2(X_t, Y_t, a^2_t) = (p_0 - p_1 X_t - p_2 Y_t) Y_t - \frac{1}{2\gamma_2} (a^2_t)^2$
    \end{itemize}
    
        where $p_0, p_1, p_2$ are positive constants.

    \item Utility Functions: Firms are assumed to have risk aversion utility functions over their total accumulated payoff:
        \begin{itemize}
            \item Firm 1: $U_1(Z) = -\exp(-\eta_1 Z)$
            \item Firm 2: $U_2(Z) = -\exp(-\eta_2 Z)$
        \end{itemize}
        where $Z = \int_0^T \pi_i(X_t, Y_t, a^i_t) dt$ is the total payoff over the horizon $[0, T]$, and $\eta_1, \eta_2 > 0$.
    \item Objective: Each firm aims to maximize its expected utility by choosing its control strategy $a^i = \{a^i_t\}_{t \in [0,T]}$. The value functions representing the maximum achievable expected utility are:
        \begin{align*}
            V_1(0,x,y) &:= \sup_{a^1} \mathbb{E}\left[ -\exp\left(-\eta_1 \int_0^T \left( \left(p_0 - p_1 X_t - p_2 Y_t\right) X_t - \frac{1}{2\gamma_1} (a^1_t)^2 \right) dt \right) \;\middle|\; X_0=x, Y_0=y \right], \\
            V_2(0,x,y) &:= \sup_{a^2} \mathbb{E}\left[ -\exp\left(-\eta_2 \int_0^T \left( \left(p_0 - p_1 X_t - p_2 Y_t\right) Y_t - \frac{1}{2\gamma_2} (a^2_t)^2 \right) dt \right) \;\middle|\; X_0=x, Y_0=y \right].
        \end{align*}

\end{itemize}

We seek a Nash Equilibrium in feedback strategies, denoted by a pair $(\hat a^1, \hat a^2)$. This pair constitutes an equilibrium if neither firm can improve its expected utility by unilaterally deviating from its equilibrium strategy, given that the other firm adheres to its equilibrium strategy. Formally, if such a pair exists, it must satisfy the following conditions for all alternative admissible strategies $a^1, a^2$:
\begin{align*}
    &\mathbb{E}\left[ U_1 \left( \int_0^T \pi_1(X_t, \hat Y_t, a^1_t) dt \right) \right]
    \leq \mathbb{E}\left[ U_1 \left( \int_0^T \pi_1(\hat X_t, \hat Y_t, \hat a^1_t) dt \right) \right], \\
    &\mathbb{E}\left[ U_2 \left( \int_0^T \pi_2(\hat X_t, Y_t, a^2_t) dt \right) \right]
    \leq \mathbb{E}\left[ U_2 \left( \int_0^T \pi_2(\hat X_t, \hat Y_t, \hat a^2_t) dt \right) \right].
\end{align*}
These expectations are evaluated considering the relevant state dynamics:
\begin{align*}
    \dd X_t &= a^1_t \, dt + \sigma_1 \, dW^1_t, \quad &&X_0 = x_0, \\
    \dd \hat X_t &= \hat a^1_t \, dt + \sigma_1 \, dW^1_t, \quad &&\hat X_0 = x_0, \\
    \dd Y_t &= a^2_t \, dt + \sigma_2 \, dW^2_t, \quad &&Y_0 = y_0, \\
    \dd \hat Y_t &= \hat a^2_t \, dt + \sigma_2 \, dW^2_t, \quad &&\hat Y_0 = y_0.
\end{align*}

\section{Results}
\subsection{Single firm with 2 technologies}
\begin{Proposition} The optimal  rates $z^*$ are given by
\begin{align*}
    &z_1^*=\frac{\gamma_1v_1+\sigma_1^2\eta_pv_1}{\sigma_1^2\eta_p+\gamma_1+\eta_A\sigma_1^2}\\
    &z_2^*=\frac{\gamma_2v_2+\sigma_2^2\eta_pv_2}{\sigma_2^2\eta_p+\gamma_2+\eta_A\sigma_2^2}
\end{align*}
with $v(t,x) =  \frac12 x \cdot A(t) x + B(t) x + C(t)$ and where $A$, $B$ and $C$ are solutions of the ODEs
\begin{align*}
0 = Q + \frac{1}{2}\dot{A} + \frac{1}{2}AMA, \quad A(T) = 0, \quad 0 = L + \dot B + AM B, \quad B(T) = 0. 
\end{align*}    
\end{Proposition}
\subsection{Two firms model}
\begin{Proposition} The incentive payments rates $z^\ast$ are given by
\begin{align}
&z_{11}^\ast = \Lambda_{12} v_1, \quad 
z_{12}^\ast = \frac{\eta_p}{\eta_1+\eta_p}(1 - \Lambda_{21}) v_2 = \frac{\eta_2\eta_p \sigma_2^2}{(\eta_1+\eta_p)(\gamma_2+(\eta_2+\bar\eta_1)\sigma_2^2)} v_2, \\
& 
\Lambda_{12} := \frac{\gamma_1 + \bar\eta_2\sigma_1^2}{\gamma_1 + (\eta_1+\bar\eta_2)\sigma_1^2}, \quad
\Lambda_{21} := \frac{\gamma_2 + \bar\eta_1\sigma_2^2}{\gamma_2 + (\eta_2+\bar\eta_1)\sigma_2^2}
\end{align}
with $v(t,x) =  \frac12 x \cdot A(t) x + B(t) x + C(t)$ and where $A$, $B$ and $C$ are solutions of the ODEs
\begin{align*}
0 = Q + \dot A + AMA, \quad A(T) = 0, \quad 0 = L + \dot B + AM B, \quad B(T) = 0. 
\end{align*}
\end{Proposition}

\subsection{Firm equilibrium without incentives}

We begin by analyzing the optimal strategy for one firm when the other firm's strategy is assumed to be a known, deterministic function of time. This corresponds to solving a standard stochastic control problem for the optimizing firm.

\begin{Proposition}[Optimal Strategy against Deterministic Competitor] \label{prop:best_response}
~ 
\begin{enumerate}
    \item Given a deterministic strategy $(a^2_t)_{t \in [0,T]}$ for Firm 2, the optimal strategy for Firm 1 is given by:
    \begin{align*}
        a^{1,\star}_t =  -\frac{\gamma_1\partial_x v_1}{\eta_1 v_1} 
    \end{align*}
    where the associated value function has the form:
    \begin{align*}
        v_1(t,x,y) = -\exp\left( \eta_1 W_1(t, x, y) \right)
    \end{align*}
    with $W_1(t,x,y)$ being the quadratic function:
    \[
        W_1(t, x, y) = \frac{1}{2} A(t) x^2 + \frac{1}{2} B(t) y^2 + C(t) xy + D(t) x + E(t) y + F(t)
    \]
    The coefficient functions $A(t), \dots, F(t)$ satisfy the following system of ODEs :
    \begin{equation}\label{eq:ode_system_best_response_1}
    \begin{cases}
        A' + \sigma_1^2 \eta_1 A^2 + \sigma_2^2 \eta_1 C^2 - 2p_1 - \gamma_1 A^2 = 0 \\
        B' + \sigma_1^2 \eta_1 C^2 + \sigma_2^2 \eta_1 B^2 - \gamma_1 C^2 = 0 \\
        C' + \sigma_1^2 \eta_1 AC + \sigma_2^2 \eta_1 BC - p_2 - \gamma_1 AC = 0 \\
        D' + \sigma_1^2 \eta_1 AD + \sigma_2^2 \eta_1 CE + a^2(t) C - \gamma_1 AD = 0 \\[1ex]
        E' + \sigma_1^2 \eta_1 CD + \sigma_2^2 \eta_1 BE + a^2(t) B - \gamma_1 CD = 0 \\[1ex]
        F' + \frac{1}{2} \sigma_1^2 (\eta_1 D^2 + A) + \frac{1}{2} \sigma_2^2 (\eta_1 E^2 + B) + a^2(t) E + p_0 - \frac{1}{2} \gamma_1 D^2 = 0
    \end{cases}
    \end{equation}
    subject to the terminal conditions $A(T) = B(T) = C(T) = D(T) = E(T) = F(T) = 0$.

    \item Similarly, given a deterministic strategy $(a^1_t)_{t \in [0,T]}$ for Firm 1, the optimal strategy for Firm 2 is given by:
    \begin{align*}
        a^{2,\star}_t = -\frac{\gamma_2\partial_y v_2}{\eta_2 v_2} 
    \end{align*}
    where the associated value function has the form:
    \begin{align*}
        v_2(t,x,y) = -\exp\left( \eta_2 W_2(t, x, y) \right)
    \end{align*}
    with $W_2(t,x,y)$ being the quadratic function:
    \[
        W_2(t, x, y) = \frac{1}{2} \tilde{A}(t) x^2 + \frac{1}{2} \tilde{B}(t) y^2 + \tilde{C}(t) xy + \tilde{D}(t) x + \tilde{E}(t) y + \tilde{F}(t)
    \]
    The coefficient functions $\tilde{A}(t), \dots, \tilde{F}(t)$ satisfy the following system of ODEs :
     \begin{equation}\label{eq:ode_system_best_response_2}
    \begin{cases}
        \tilde{A}' + \sigma_1^2 \eta_2 \tilde{A}^2 + \sigma_2^2 \eta_2 \tilde{C}^2 - \gamma_2 \tilde{C}^2 = 0 \\
        \tilde{B}' + \sigma_1^2 \eta_2 \tilde{C}^2 + \sigma_2^2 \eta_2 \tilde{B}^2 - 2p_2 - \gamma_2 \tilde{B}^2 = 0 \\
        \tilde{C}' + \sigma_1^2 \eta_2 \tilde{A}\tilde{C} + \sigma_2^2 \eta_2 \tilde{B}\tilde{C} - p_1 - \gamma_2 \tilde{B}\tilde{C} = 0 \\
        \tilde{D}' + \sigma_1^2 \eta_2 \tilde{A}\tilde{D} + \sigma_2^2 \eta_2 \tilde{C}\tilde{E} + a^1(t) \tilde{A} - \gamma_2 \tilde{C}\tilde{E} = 0 \\[1ex]
        \tilde{E}' + \sigma_1^2 \eta_2 \tilde{C}\tilde{D} + \sigma_2^2 \eta_2 \tilde{B}\tilde{E} + a^1(t) \tilde{C} - \gamma_2 \tilde{B}\tilde{E} = 0 \\[1ex]
        \tilde{F}' + \frac{1}{2} \sigma_1^2 (\eta_2 \tilde{D}^2 + \tilde{A}) + \frac{1}{2} \sigma_2^2 (\eta_2 \tilde{E}^2 + \tilde{B}) + a^1(t) \tilde{D} + p_0 - \frac{1}{2} \gamma_2 \tilde{E}^2 = 0
    \end{cases}
    \end{equation}
    subject to the terminal conditions $\tilde{A}(T) = \tilde{B}(T) = \tilde{C}(T) = \tilde{D}(T) = \tilde{E}(T) = \tilde{F}(T) = 0$.
\end{enumerate}
\end{Proposition}

\begin{figure}[H]
    \centering
    \includegraphics[width=0.8\textwidth]{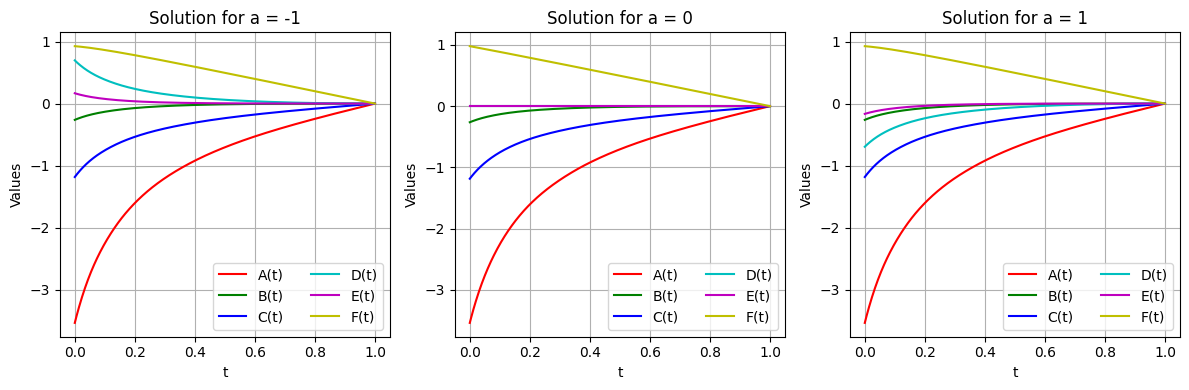} 
    \caption[Optimal Coefficients for Firm 1 vs Constant $a^2$]{Numerical solution for Firm 1's coefficients $A(t), \dots, F(t)$ when Firm 2 employs different constant strategies $a^2$.}
    \label{fig:best_response_solution}
    \vspace{1em} 
    \begin{minipage}{0.8\textwidth} 
    \small 
    \textit{Parameters used:}
    \begin{itemize} 
        \item $\sigma_1 = 0.2$, $\sigma_2 = 0.3$
        \item $p_0 = 1.0$, $p_1 = 0.6$, $p_2 = 0.4$
        \item $\eta_1 = 1.0$, $\eta_2 = 1.0$
        \item $\gamma_1 = 1.5$, $\gamma_2 = 1.0$
    \end{itemize}
    \end{minipage}
\end{figure}

\paragraph{Numerical Illustration and Interpretation}
Solving the system of ODEs \eqref{eq:ode_system_best_response_1} numerically allows us to understand how Firm 1's optimal strategy coefficients evolve, given a specific path for $a^2(t)$. Figure~\ref{fig:best_response_solution} illustrates the solution for different constant values of $a^2$.

While a detailed economic interpretation requires domain expertise (which is out of reach for us), the plots suggest that the competitor's strategy $a^2$ primarily influences the coefficients $D(t)$ and $E(t)$. Since Firm 1's optimal strategy is $a^{1,\star}_t = -\gamma_1[A(t)x+C(t)y+D(t)]$, the dependence of $D(t)$ on $a^2$ directly affects Firm 1's optimal actions.

\paragraph{Nash Equilibrium Analysis}

The more complex scenario involves finding the Nash equilibrium, where both firms simultaneously choose their optimal strategies, taking the other's optimal strategy into account. This leads to a coupled system where each firm's optimization depends on the outcome of the other's.

\begin{Proposition}[Nash Equilibrium Characterization] \label{prop:nash_equilibrium}
If a Nash equilibrium $(\hat a^1, \hat a^2)$ exists then:
\begin{enumerate}
    \item The equilibrium strategies are given by:
    \begin{align*}
        \hat a^{1}_t = -\frac{\gamma_1\partial_x \hat v_1}{\eta_1 \hat v_1} \\
        \hat a^{2}_t = -\frac{\gamma_2\partial_y \hat v_2}{\eta_2 \hat v_2}
    \end{align*}
    \item The associated value functions are:
    \begin{align*}
        \hat v_1(t,x,y) &= -\exp\left( \eta_1 W_1(t, x, y) \right)\\
        \hat v_2(t,x,y) &= -\exp\left( \eta_2 W_2(t, x, y) \right)
    \end{align*}
    where $W_1$ and $W_2$ are the quadratic functions defined previously.
    \item The coefficients $(A(t), \dots, F(t))$ and $(\tilde{A}(t), \dots, \tilde{F}(t))$ satisfy the following coupled system of 12 ODEs:
\end{enumerate}
\end{Proposition}

\begin{equation}\label{eq:ode_system_nash_1}
\begin{cases}
A' + \sigma_1^2 \eta_1 A^2 + \sigma_2^2 \eta_1 C^2 - 2\gamma_2 \tilde{C} C - 2p_1 - \gamma_1 A^2 = 0 \\[1ex]
B' + \sigma_1^2 \eta_1 C^2 + \sigma_2^2 \eta_1 B^2 - 2\gamma_2 \tilde{B} B - \gamma_1 C^2 = 0 \\[1ex]
C' + \sigma_1^2 \eta_1 AC + \sigma_2^2 \eta_1 BC - \gamma_2 (\tilde{B}C + \tilde{C}B) - p_2 - \gamma_1 AC = 0 \\[1ex]
D' + \sigma_1^2 \eta_1 AD + \sigma_2^2 \eta_1 CE - \gamma_2 (\tilde{C}E + \tilde{E}C) - \gamma_1 AD = 0 \\[1ex]
E' + \sigma_1^2 \eta_1 CD + \sigma_2^2 \eta_1 BE - \gamma_2 (\tilde{B}E + \tilde{E}B) - \gamma_1 CD = 0 \\[1ex]
F' + \frac{1}{2} \sigma_1^2 (\eta_1 D^2 + A) + \frac{1}{2} \sigma_2^2 (\eta_1 E^2 + B) - \gamma_2 \tilde{E} E + p_0 - \frac{1}{2} \gamma_1 D^2 = 0
\end{cases}
\end{equation}


\begin{equation}\label{eq:ode_system_nash_2}
\begin{cases}
\tilde{A}' + \sigma_1^2 \eta_2 \tilde{A}^2 + \sigma_2^2 \eta_2 \tilde{C}^2 - 2\gamma_1 A \tilde{A} - \gamma_2 \tilde{C}^2 = 0 \\[1ex]
\tilde{B}' + \sigma_1^2 \eta_2 \tilde{C}^2 + \sigma_2^2 \eta_2 \tilde{B}^2 - 2\gamma_1 C \tilde{C} - 2p_2 - \gamma_2 \tilde{B}^2 = 0 \\[1ex]
\tilde{C}' + \sigma_1^2 \eta_2 \tilde{A}\tilde{C} + \sigma_2^2 \eta_2 \tilde{B}\tilde{C} - \gamma_1 (A\tilde{C} + C\tilde{A}) - p_1 - \gamma_2 \tilde{B}\tilde{C} = 0 \\[1ex]
\tilde{D}' + \sigma_1^2 \eta_2 \tilde{A}\tilde{D} + \sigma_2^2 \eta_2 \tilde{C}\tilde{E} - \gamma_1 (A\tilde{D} + D\tilde{A}) - \gamma_2 \tilde{C}\tilde{E} = 0 \\[1ex]
\tilde{E}' + \sigma_1^2 \eta_2 \tilde{C}\tilde{D} + \sigma_2^2 \eta_2 \tilde{B}\tilde{E} - \gamma_1 (C\tilde{D} + D\tilde{C}) - \gamma_2 \tilde{B}\tilde{E} = 0 \\[1ex]
\tilde{F}' + \frac{1}{2} \sigma_1^2 (\eta_2 \tilde{D}^2 + \tilde{A}) + \frac{1}{2} \sigma_2^2 (\eta_2 \tilde{E}^2 + \tilde{B}) - \gamma_1 D\tilde{D} + p_0 - \frac{1}{2} \gamma_2 \tilde{E}^2 = 0
\end{cases}
\end{equation}

Both systems must be solved subject to the terminal conditions at $t=T$:
\begin{gather*}
A(T) = B(T) = C(T) = D(T) = E(T) = F(T) = 0 \\
\tilde{A}(T) = \tilde{B}(T) = \tilde{C}(T) = \tilde{D}(T) = \tilde{E}(T) = \tilde{F}(T) = 0
\end{gather*}

\begin{figure}[H] 
    \centering
    \includegraphics[width=0.8\textwidth]{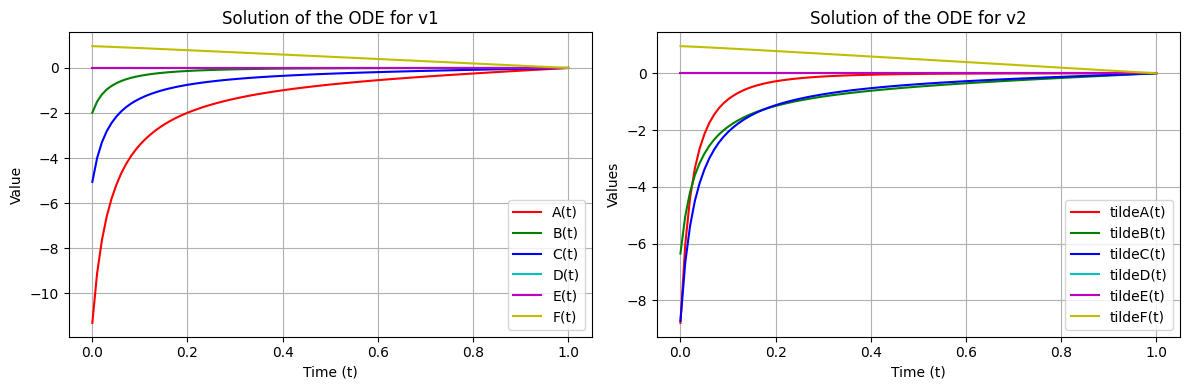} 
    \caption[Nash Equilibrium Coefficients]{Numerical solution of the coupled ODE systems \eqref{eq:ode_system_nash_1} and \eqref{eq:ode_system_nash_2}.}
    \label{fig:nash_equilibrium_solution}
    \vspace{1em}
    \begin{minipage}{0.8\textwidth}
    \small
    \textit{Parameters used:}
    \begin{itemize} 
        \item $\sigma_1 = 0.2$, $\sigma_2 = 0.3$
        \item $p_0 = 1.0$, $p_1 = 0.6$, $p_2 = 0.4$
        \item $\eta_1 = 1.0$, $\eta_2 = 1.0$
        \item $\gamma_1 = 1.5$, $\gamma_2 = 1.0$
    \end{itemize}
    \end{minipage}
\end{figure}

\paragraph{Numerical Resolution}
Due to the coupled and non-linear nature system, we will use numerical methods for solution. Figure~\ref{fig:nash_equilibrium_solution} shows an example solution.

\section{Conclusions}
This study demonstrates how stochastic control techniques can be used to model and analyze decarbonization strategies in the energy market. In the single-firm case, the transition to clean technology depends mainly on cost differentials and market uncertainty. When competition is introduced, strategic interactions can delay or prevent optimal decarbonization, highlighting a market failure. The introduction of incentives proves to be a powerful tool in aligning private decisions with social objectives.

Our results emphasize the need for well-designed regulatory interventions to support clean energy investments, particularly in competitive markets. Future research could extend these models by incorporating learning effects, heterogeneous firms, or more complex regulatory instruments. Nonetheless, our work provides a foundational step toward understanding and guiding the decarbonization of energy systems under uncertainty.
\appendix

\section{Single firm with 2 technologies}
\paragraph{Proof}
The PDE satisfied by V is given by
\begin{align*}
    &0=\partial_t V +\frac{1}{2}(\sigma_1^2 V_{11}+\sigma_2^2 V_{22})-f(x)V_y+\eta_p g(x)V+
    \sup_Z \tilde{h}(z,x,Dv)\\
    &\tilde{h}(z,x,Dv)=\frac{1}{2}(\sigma_1^2 z_1^2 +\sigma_2^2 z_2^2)V_{yy}+\sigma_1^2 z_1 V_{1y}+\sigma_2^2 z_2 V_{2y}\\
    &+V_1 \gamma_1 z_1 +V_2\gamma_2 z_2 +V_y (c(\hat{a}(z))+\frac{1}{2}\eta_A(\sigma_1^2 z_1^2 +\sigma_2^2 z_2^2) ]
\end{align*}
Now, let's consider that $V(t,x,y)=U_p(v(t,x)-y)$, we obtain
\begin{align*}
    &\partial_t V =-\eta_p V\partial_t v, \, V_1 = -\eta_p V v_1, \, V_2 = -\eta_p V v_2, \, V_y = \eta_p V, \\
& V_{ii} = \eta_p^2 v_i^2V- \eta_p V v_{ii}, \, V_{yy} = \eta_p^2 V, \, V_{iy} = - \eta_p^2 V v_i,i=1,2.
\end{align*}
After that we get
\begin{align*}
    &0=\partial_t v+\frac{1}{2}\sigma_1^2(v_{11}-\eta_p v_1^2)+\frac{1}{2}\sigma_2^2(v_{22}-\eta_pv_2^2)+f(x)-g(x)+\sup_z h(z,x,Dv)\\
    &v(T,x)=0\\
&h(z,x,Dv)=\gamma_1 z_1v_1+\gamma_2z_2v_2 +\sigma_1^2z_1\eta_p v_1+\sigma_2^2z_2\eta_pv_2-\frac{1}{2}(\sigma_1^2z_1^2+\sigma_2^2z_2^2)\eta_p-c(\hat{a}(z))-\frac{1}{2}\eta_A(\sigma_1^2z_1^2+\sigma_2^2z_2^2)
\end{align*}
Then with $c(a)=\frac{1}{2}\frac{a_1^2}{\gamma_1}+\frac{1}{2}\frac{a_2^2}{\gamma_2}$, we have
\begin{align*}
    h(z,x,Dv)&=\gamma_1 z_1v_1+\gamma_2z_2v_2 +\sigma_1^2z_1\eta_p v_1+\sigma_2^2z_2\eta_pv_2-\frac{1}{2}(\sigma_1^2z_1^2+\sigma_2^2z_2^2)\eta_p-\frac{1}{2}\gamma_1 z_1^2-\frac{1}{2}\gamma_2 z_2^2-\frac{1}{2}\eta_A(\sigma_1^2z_1^2+\sigma_2^2z_2^2)\\
    &=(\gamma_1v_1+\sigma_1^2\eta_pv_1)z_1-\frac{1}{2}(\sigma_1^2\eta_p+\gamma_1+\eta_A\sigma_1^2)z_1^2\\
    &+(\gamma_2v_2+\sigma_2^2\eta_pv_2)z_2-\frac{1}{2}(\sigma_2^2\eta_p+\gamma_2+\eta_A\sigma_2^2)z_2^2
\end{align*}
Hence the optimal value of $z_1^*$ and $z_2^*$ are
\begin{align*}
    &z_1^*=\frac{\gamma_1v_1+\sigma_1^2\eta_pv_1}{\sigma_1^2\eta_p+\gamma_1+\eta_A\sigma_1^2}\\
    &z_2^*=\frac{\gamma_2v_2+\sigma_2^2\eta_pv_2}{\sigma_2^2\eta_p+\gamma_2+\eta_A\sigma_2^2}
\end{align*}
Finally the PDE of $v$ becomes
\begin{align*}
    &0=\partial_t v +f(x)-g(x) \\
    &+\frac{1}{2}\sigma_1^2v_{11}+(\frac{(\gamma_1+\sigma_1^2\eta_p)^2}{\sigma_1^2\eta_p+\gamma_1+\eta_A\sigma_1^2}-\frac{1}{2}\eta_p)v_1^2\\
    &+\frac{1}{2}\sigma_2^2v_{22}+(\frac{(\gamma_2+\sigma_2^2\eta_p)^2}{\sigma_2^2\eta_p+\gamma_2+\eta_A\sigma_2^2}-\frac{1}{2}\eta_p)v_2^2\\
    &0=v(T,x)
\end{align*}
As we have seen in class, this PDE has the following form
\begin{align*}
    &0=Lx+\frac{1}{2}x\cdot Qx +\partial_t v +\frac{1}{2}Tr(\Sigma\Sigma^TD^2v)+\frac{1}{2}Dv\cdot M Dv\\
    &0=v(T,x)
\end{align*}
Assume that $v(t,x)=\frac{1}{2}x\cdot Ax+Bx+C$. Where $A$ is symmetric and depends on time, $B$ and $C$ depends on time. Then, $Dv=Ax+B$ and $Dv\cdot M Dv=(Ax+B)\cdot (MAx+MB)$ and a system of ODE
\begin{align*}
    &0=Q+\frac{1}{2}\dot{A}+\frac{1}{2}AMA\\
    &0=A(T)\\
    &0=L+\dot{B}+AMB\\
    &0=B(T)
\end{align*}
\section{Two firms model}

The PDE satisfied by $V$ is given by
\begin{align*}
&0 = \partial_t V + \frac12 \sigma_1^2 V_{11} + \frac12 \sigma_2^2 V_{22} - f(x) V_y +\eta_p g(x) V
+  \sup_{z} \tilde h(z,x,Dv), \quad V(T,x,y) = U_{\rm P}(-y), \\
& \tilde h(z,x,Dv) := \gamma_1 z_{11} V_1 + \gamma_2 z_{22} V_2 + \big[c(\hat a(z)) + R(z)\big] V_y 
+ \frac12 \big(\hat\sigma_1^2(z) + \hat\sigma_2^2(z)\big) V_{yy}
+ \sigma_1 \hat\sigma_1(z) V_{1y} + \sigma_2 \hat\sigma_2(z) V_{2y} 
\end{align*}
Now, consider $V(t,x,y) := U_{\rm P}(v(t,x)- y)$, we have
\begin{align*}
&\partial_t V = -\eta_p V  \partial_t v, \, V_1 = -\eta_p V v_1, \, V_2 = -\eta_p V v_2, \, V_y = \eta_p V, \\
& V_{ii} = -\eta_p V(v_{ii} - \eta_p v_i^2), \, V_{yy} = \eta_p^2 V, \, V_{iy} = - \eta_p^2 V v_i.
\end{align*}
Noting that $-\eta_p V>0$, we have
\begin{align*}
&0 = \partial_t v + \frac12 \sigma_1^2 (v_{11}-\eta_p v_1^2) + \frac12 \sigma_2^2 (v_{22}-\eta_p v_2^2) + f(x) -  g(x) +  \sup_{z} h(z,x,Dv), \quad v(T,x) = 0, \\
& h(z,x,Dv) := \gamma_1 z_{11} v_1 + \gamma_2 z_{22} v_2 - \big[c(\hat a(z)) + R(z)\big]  
- \frac12 \eta_p \big(\hat\sigma_1^2(z) + \hat\sigma_2^2(z)\big)  
+ \sigma_1 \eta_p \hat\sigma_1(z) v_1 
+ \sigma_2 \eta_p \hat\sigma_2(z) v_2 
\end{align*}
Now focus on $h$
\begin{align*}
 h(z,x,Dv) & = \gamma_1 z_{11} v_1 + \gamma_2 z_{22} v_2 
 - \frac12 \big[ \gamma_1 z_{11}^2 + \gamma_2 z_{22}^2  + 
  \eta_1 \sigma_1^2 z_{11}^2 + \eta_1 \sigma_2^2 z_{12}^2 + \eta_2 \sigma_2^2 z_{22}^2 + \eta_2 \sigma_1^2 z_{21}^2\big]  \\
& - \frac12 \eta_p \big(\sigma_1^2(z_{11}+z_{21})^2 + \sigma_2^2(z_{22}+z_{12})^2\big)  
+  \eta_p \sigma_1^2(z_{11}+z_{21}) v_1 
+  \eta_p  \sigma_2^2(z_{22}+z_{12}) v_2, \\
& = 
(\gamma_1+\eta_p\sigma_1^2) v_1 z_{11} 
+ (\gamma_2+\eta_p\sigma_2^2) v_2 z_{22} 
- \frac12\big[ (\gamma_1+(\eta_p+\eta_1)\sigma_1^2) z_{11}^2  + (\gamma_2+(\eta_p + \eta_2)\sigma_2^2) z_{22}^2\big] \\
& -\frac12 (\eta_1+\eta_p) \sigma_2^2 z_{12}^2 -\frac12 (\eta_2+\eta_p) \sigma_1^2 z_{21}^2 
- \eta_p \sigma_1^2 z_{11} z_{21} - \eta_p \sigma_2^2 z_{22} z_{12}
+ \eta_p \sigma_2^2 v_2 z_{12}
+ \eta_p \sigma_1^2 v_1 z_{21}, \\
& =
(\gamma_1+\eta_p\sigma_1^2) v_1 z_{11} 
+ (\gamma_2+\eta_p\sigma_2^2) v_2 z_{22} 
- \frac12\big[ (\gamma_1+(\eta_p+\eta_1)\sigma_1^2) z_{11}^2  + (\gamma_2+(\eta_p + \eta_2)\sigma_2^2) z_{22}^2\big] \\
& -\frac12 (\eta_1+\eta_p) \sigma_2^2 \Big[ z_{12}
+ \frac{\eta_p}{\eta_1+\eta_p}(z_{22}-v_2)\Big]^2 
+ \frac12 \frac{\eta_p^2  \sigma_2^2}{\eta_1+\eta_p} (z_{22}-v_2)^2 \\
& -\frac12 (\eta_2+\eta_p) \sigma_1^2 \Big[z_{21}
+ \frac{\eta_p}{\eta_2+\eta_p}(z_{11}-v_1)\Big]^2 
+ \frac12 \frac{\eta_p^2\sigma_1^2 }{\eta_2+\eta_p}  (z_{11}-v_1)^2, \\
\end{align*}
Hence $z_{12} = \eta_{1p} (v_2 - z_{22})$ and $z_{21} = \eta_{2p}(v_1 - z_{11})$, with $\eta_{ip} := \frac{\eta_p}{\eta_i+\eta_p}$, and we are left with
\begin{align*}
 h(z,x,Dv) & = 
 \Big[ \gamma_1 + \Big(1  - \frac{\eta_p}{\eta_2+\eta_p}\Big) \eta_p \sigma_1^2 \Big] v_1 z_{11}
- \frac12 \Big[ \gamma_1+\Big\{ (\eta_p+\eta_1) - \frac{\eta_p^2}{\eta_2+\eta_p}\Big\} \sigma_1^2 \Big] z_{11}^2\\
 & 
+ 
 \Big[ \gamma_2 + \Big(1  - \frac{\eta_p}{\eta_1+\eta_p}\Big) \eta_p \sigma_2^2 \Big] v_2 z_{22} 
- \frac12 \Big[ \gamma_2+\Big\{ (\eta_p+\eta_2) - \frac{\eta_p^2}{\eta_1+\eta_p}\Big\} \sigma_2^2 \Big] z_{22}^2\\
& 
+ \frac12 \frac{\eta_p^2  \sigma_1^2}{\eta_2+\eta_p} v_1^2
+  \frac12 \frac{\eta_p^2  \sigma_2^2}{\eta_1+\eta_p} v_2^2 
\end{align*}
Hence with $\bar\eta_i := \frac{\eta_p\eta_i}{\eta_p+\eta_i}$, 
\begin{align*}
 h(z,x,Dv) & = 
 \Big[ \gamma_1 +  \bar\eta_2 \sigma_1^2 \Big] v_1 z_{11}
- \frac12 \Big[ \gamma_1+ (\bar\eta_2+\eta_1)\sigma_1^2 \Big] z_{11}^2
 + 
 \Big[ \gamma_2 +  \bar\eta_1 \sigma_2^2 \Big] v_2 z_{22} 
- \frac12 \Big[ \gamma_2+ (\bar\eta_1+\eta_2)\sigma_2^2 \Big] z_{22}^2\\
& 
+ \frac12 \frac{\eta_p^2  \sigma_1^2}{\eta_2+\eta_p} v_1^2
+ \frac12 \frac{\eta_p^2  \sigma_2^2}{\eta_1+\eta_p} v_2^2 \\
 = &
- \frac12 \Big[ \gamma_1+ (\bar\eta_2+\eta_1)\sigma_1^2 \Big]
 \Big( z_{11} - \frac{\gamma_1 + \bar\eta_2\sigma_1^2}{\gamma_1 + (\eta_1+\bar\eta_2)\sigma_1 ^2} v_1\Big)^2 
 + \frac12 \Big[\frac{(\gamma_1 + \bar\eta_2\sigma_1^2)^2}{\gamma_1 + (\eta_1+\bar\eta_2)\sigma_1 ^2} + \frac{\eta_p^2  \sigma_1^2}{\eta_2+\eta_p} \Big]v_1^2 \\
&- \frac12 \Big[ \gamma_2 + (\bar\eta_1+\eta_2)\sigma_2^2 \Big]
 \Big( z_{22} - \frac{\gamma_2 + \bar\eta_1\sigma_2^2}{\gamma_2 + (\eta_2+\bar\eta_1)\sigma_2^2} v_2\Big)^2 
 + \frac12 \Big[\frac{(\gamma_2 + \bar\eta_1\sigma_2^2)^2}{\gamma_2 + (\eta_2+\bar\eta_1)\sigma_2^2} + \frac{\eta_p^2  \sigma_2^2}{\eta_1+\eta_p} \Big]v_2^2
\end{align*}
Hence, we have
\begin{align*}
&z_{11}^\ast = \Lambda_{12} v_1, \quad 
z_{12}^\ast = \frac{\eta_p}{\eta_1+\eta_p}(1 - \Lambda_{21}) v_2 = \frac{\eta_2\eta_p \sigma_2^2}{(\eta_1+\eta_p)(\gamma_2+(\eta_2+\bar\eta_1)\sigma_2^2)} v_2, \\
& 
\Lambda_{12} := \frac{\gamma_1 + \bar\eta_2\sigma_1^2}{\gamma_1 + (\eta_1+\bar\eta_2)\sigma_1^2}, \quad
\Lambda_{21} := \frac{\gamma_2 + \bar\eta_1\sigma_2^2}{\gamma_2 + (\eta_2+\bar\eta_1)\sigma_2^2}
\end{align*}
And the PDE for $v$ becomes
\begin{align*}
&0 =  f(x) -  g(x) + \partial_t v + \frac12 \sigma_1^2 v_{11}  + \frac12 \sigma_2^2 v_{22} 
+  \frac12\Big[\frac{(\gamma_1 + \bar\eta_2\sigma_1^2)^2}{\gamma_1 + (\eta_1+\bar\eta_2)\sigma_1^2} + \bar\eta_2\sigma_1^2\Big] v_1^2 
+  \frac12\Big[\frac{(\gamma_2 + \bar\eta_1\sigma_2^2)^2}{\gamma_2 + (\eta_2+\bar\eta_1)\sigma_2^2} + \bar\eta_1\sigma_2^2\Big] v_2^2 
  , \quad v(T,x) = 0.
\end{align*}
This PDE takes the form 
\begin{align*}
&0 =   L x + \frac12 x \cdot Q x + \partial_t v +  \frac12 {\rm Tr}[\Sigma\Sigma^\intercal D ^2v] + \frac12 Dv \cdot M Dv
  , \quad v(T,x) = 0, 
\end{align*}
Assume $v(t,x) = \frac12 x\cdot Ax + Bx + C$ with $A, B$ and $C$ functions of time and $A$ symmetric, we get the ODEs for $A$ and $B$ as
\begin{align*}
0 = Q + \dot A + AMA, \quad A(T) = 0, \quad 0 = L + \dot B + AM B, \quad B(T) = 0. 
\end{align*}

\section{Firm without incentives}

\paragraph{Proof}
To characterize the optimal strategy $\hat a^1$ for Firm 1, given Firm 2's equilibrium strategy $\hat a^2$, we solve the value function with a Hamilton-Jacobi-Bellman equation. As done in class, we will admit this equation :
\begin{align*}
\begin{cases}
    0 = \partial_t V_1 + \frac{1}{2} \sigma_1^2 \partial_{xx}V_1 + \frac{1}{2} \sigma_2^2 \partial_{yy}V_1 
    + \hat a^2_t \partial_y V_1 
    + \eta_1 \left( p_0 - p_1 x^2 - p_2 y x \right) V_1 
 + \inf_{a^1 \in \mathbb{R}} \left\{ \partial_x V_1 a^1 + \frac{\eta_1 V_1}{2\gamma_1} (a^1)^2 \right\} \quad &\forall (t,x,y), \\
    V_1(T,x,y) = -1 \quad &\forall (x,y).
\end{cases}
\end{align*}

Solving the minimum : $\hat a^1 =  - \frac{\gamma_1}{\eta_1 V_1} \partial_x V_1$ and we obtain the PDE :

\begin{align*}
\begin{cases}
0 = \partial_t V_1 + \frac{1}{2} \sigma_1^2 \partial_{xx}V_1 + \frac{1}{2} \sigma_2^2 \partial_{yy}V_1 + \hat a^2_t \partial_y V_1 + \eta_1 \left( p_0 - p_1 x^2 - p_2 y x \right) V_1 -  \frac{\gamma_1}{2\eta_1 V_1} (\partial_x V_1)^2 &\forall (t,x,y), \\
    V_1(T,x,y) = -1 \quad &\forall (x,y).
\end{cases}
\end{align*} \\

\noindent
We propose the following ansatz for $V_1$:
\begin{align*}
V_1(t, x, y) = -\exp \left( \eta_1 W_1(t, x, y) \right) 
\end{align*}
where $W_1(t, x, y)$ is a quadratic function of the state variables:
\begin{align*}
    W_1(t, x, y) = \frac{1}{2} A(t) x^2 + \frac{1}{2} B(t) y^2 + C(t) xy + D(t) x + E(t) y + F(t) 
\end{align*}
\noindent
$A(t), B(t), C(t), D(t), E(t), F(t)$ are deterministic functions of time, satisfying the terminal conditions $A(T)=B(T)=C(T)=D(T)=E(T)=F(T)=0$.\\

\noindent
Partial derivatives of $V_1$ :
\begin{align*}
\partial_t V_1 &= V_1 \eta_1 \partial_t W_1 = V_1\eta_1\left[ \frac{1}{2} A' x^2 + \frac{1}{2} B' y^2 + C' xy + D' x + E' y + F' \right]\\
\partial_x V_1 &= V_1 \eta_1 \partial_x W_1 = V_1 \eta_1 [A(t) x + C(t) y + D(t)] \\
-\frac{\gamma_1}{2\eta_1 V_1} (\partial_x V_1)^2 &=  - \frac{1}{2} \gamma_1 \eta_1 V_1 [A(t) x + C(t) y + D(t)]^2\\
\partial_y V_1 &= V_1 \eta_1 \partial_y W_1 = V_1 \eta_1 [B(t) y + C(t) x + E(t)] \\
\partial_{xx} V_1
&= V_1 \eta_1\left( \eta_1 [A(t) x + C(t) y + D(t)]^2 +  A(t) \right) \\
\partial_{yy} V_1 
&= V_1 \eta_1\left( \eta_1 [B(t) y + C(t) x + E(t)]^2 +  B(t) \right)\\
\end{align*}

\noindent
By replacing and dividing by $V_1\eta_1$ we obtain the following PDE : 

\begin{align*}
0 = \; &  \frac{1}{2} A' x^2 + \frac{1}{2} B' y^2 + C' xy + D' x + E' y + F'  \\
& + \frac{1}{2} \sigma_1^2 \left( \eta_1 [A x + C y + D]^2 + A \right) \\
& + \frac{1}{2} \sigma_2^2 \left( \eta_1 [B y + C x + E]^2 + B \right) \\
& + \hat a^2 (B y + C x + E) \\
& + \left( p_0 - p_1 x^2 - p_2 y x \right) \\
& - \frac{1}{2} \gamma_1 [A x + C y + D]^2\\
\end{align*}

\noindent
Following a similar method for Firm 2, we have : 

\begin{align*}
\begin{cases}
0 = \partial_t V_2 + \frac{1}{2} \sigma_1^2 \partial_{xx}V_2 + \frac{1}{2} \sigma_2^2 \partial_{yy}V_2 + \hat{a}^1_t \partial_x V_2 + \eta_2 \left( p_0 - p_2 y^2 - p_1 y x \right) V_2 -  \frac{\gamma_2}{2\eta_2 V_2} (\partial_y V_2)^2 &\forall (t,x,y), \\
    V_2(T,x,y) = 1 \quad &\forall (x,y).\\
\end{cases}
\end{align*}

\noindent
Using the ansatz $V_2(t, x, y) = \exp(\eta_2 W_2(t, x, y))$, where
\[
W_2(t, x, y) = \frac{1}{2} \tilde{A}(t) x^2 + \frac{1}{2} \tilde{B}(t) y^2 + \tilde{C}(t) xy + \tilde{D}(t) x + \tilde{E}(t) y + \tilde{F}(t)
\]
and substituting into the optimized HJB equation for $V_2$, we obtain the following PDE for $W_2(t, x, y)$:

\begin{align*}
0 = \; & \frac{1}{2} \tilde{A}' x^2 + \frac{1}{2} \tilde{B}' y^2 + \tilde{C}' xy + \tilde{D}' x + \tilde{E}' y + \tilde{F}'  \\
& + \frac{1}{2} \sigma_1^2 \left( \eta_2 [\tilde{A} x + \tilde{C} y + \tilde{D}]^2 + \tilde{A} \right) \\
& + \frac{1}{2} \sigma_2^2 \left( \eta_2 [\tilde{B} y + \tilde{C} x + \tilde{E}]^2 + \tilde{B} \right) \\
& +\hat{a}^1 (\tilde{A} x + \tilde{C} y + \tilde{D}) \\
& + \left( p_0 - p_2 y^2 - p_1 y x \right) \\
& - \frac{1}{2} \gamma_2 [\tilde{B} y + \tilde{C} x + \tilde{E}]^2\\
\end{align*}

\noindent
Now, by writing : 
\begin{align*}
\hat{a}^1_t &= - \gamma_1 (A(t) x + C(t) y + D(t))  \\
\hat{a}^2_t &= - \gamma_2 (\tilde{B}(t) y + \tilde{C}(t) x + \tilde{E}(t)) \\
\end{align*}

\noindent
We obtain :

\begin{align*}
\begin{cases}
0 =& \frac{1}{2} A' x^2 + \frac{1}{2} B' y^2 + C' xy + D' x + E' y + F' + \frac{1}{2} \sigma_1^2 \left( \eta_1 [A x + C y + D]^2 + A \right) + \frac{1}{2} \sigma_2^2 \left( \eta_1 [B y + C x + E]^2 + B \right) \\
& - \gamma_2 (\tilde{B} y + \tilde{C} x + \tilde{E}) (B y + C x + E)  + \left( p_0 - p_1 x^2 - p_2 y x \right)  - \frac{1}{2} \gamma_1 [A x + C y + D]^2\\
0 =& \frac{1}{2} \tilde{A}' x^2 + \frac{1}{2} \tilde{B}' y^2 + \tilde{C}' xy + \tilde{D}' x + \tilde{E}' y + \tilde{F}'+ \frac{1}{2} \sigma_1^2 \left( \eta_2 [\tilde{A} x + \tilde{C} y + \tilde{D}]^2 + \tilde{A} \right) + \frac{1}{2} \sigma_2^2 \left( \eta_2 [\tilde{B} y + \tilde{C} x + \tilde{E}]^2 + \tilde{B} \right) \\
& - \gamma_1 (Ax + Cy + D) (\tilde{A} x + \tilde{C} y + \tilde{D})  + \left( p_0 - p_2 y^2 - p_1 y x \right) - \frac{1}{2} \gamma_2 [\tilde{B} y + \tilde{C} x + \tilde{E}]^2\\ \end{cases}
\end{align*}

\noindent
Equating the coefficients of $x^2, y^2, xy, x, y,$ and the constant term to zero in the PDE gives the following 2 system of ODEs :

\begin{align*}
\begin{cases}
A' + \sigma_1^2 \eta_1 A^2 + \sigma_2^2 \eta_1 C^2 - 2\gamma_2 \tilde{C} C - 2p_1 - \gamma_1 A^2 = 0 \\[1ex]
B' + \sigma_1^2 \eta_1 C^2 + \sigma_2^2 \eta_1 B^2 - 2\gamma_2 \tilde{B} B - \gamma_1 C^2 = 0 \\[1ex]
C' + \sigma_1^2 \eta_1 AC + \sigma_2^2 \eta_1 BC - \gamma_2 (\tilde{B}C + \tilde{C}B) - p_2 - \gamma_1 AC = 0 \\[1ex]
D' + \sigma_1^2 \eta_1 AD + \sigma_2^2 \eta_1 CE - \gamma_2 (\tilde{C}E + \tilde{E}C) - \gamma_1 AD = 0 \\[1ex]
E' + \sigma_1^2 \eta_1 CD + \sigma_2^2 \eta_1 BE - \gamma_2 (\tilde{B}E + \tilde{E}B) - \gamma_1 CD = 0 \\[1ex]
F' + \frac{1}{2} \sigma_1^2 (\eta_1 D^2 + A) + \frac{1}{2} \sigma_2^2 (\eta_1 E^2 + B) - \gamma_2 \tilde{E} E + p_0 - \frac{1}{2} \gamma_1 D^2 = 0
\end{cases}
\end{align*}

\begin{align*}
\begin{cases}
\tilde{A}' + \sigma_1^2 \eta_2 \tilde{A}^2 + \sigma_2^2 \eta_2 \tilde{C}^2 - 2\gamma_1 A \tilde{A} - \gamma_2 \tilde{C}^2 = 0 \\[1ex]
\tilde{B}' + \sigma_1^2 \eta_2 \tilde{C}^2 + \sigma_2^2 \eta_2 \tilde{B}^2 - 2\gamma_1 C \tilde{C} - 2p_2 - \gamma_2 \tilde{B}^2 = 0 \\[1ex]
\tilde{C}' + \sigma_1^2 \eta_2 \tilde{A}\tilde{C} + \sigma_2^2 \eta_2 \tilde{B}\tilde{C} - \gamma_1 (A\tilde{C} + C\tilde{A}) - p_1 - \gamma_2 \tilde{B}\tilde{C} = 0 \\[1ex]
\tilde{D}' + \sigma_1^2 \eta_2 \tilde{A}\tilde{D} + \sigma_2^2 \eta_2 \tilde{C}\tilde{E} - \gamma_1 (A\tilde{D} + D\tilde{A}) - \gamma_2 \tilde{C}\tilde{E} = 0 \\[1ex]
\tilde{E}' + \sigma_1^2 \eta_2 \tilde{C}\tilde{D} + \sigma_2^2 \eta_2 \tilde{B}\tilde{E} - \gamma_1 (C\tilde{D} + D\tilde{C}) - \gamma_2 \tilde{B}\tilde{E} = 0 \\[1ex]
\tilde{F}' + \frac{1}{2} \sigma_1^2 (\eta_2 \tilde{D}^2 + \tilde{A}) + \frac{1}{2} \sigma_2^2 (\eta_2 \tilde{E}^2 + \tilde{B}) - \gamma_1 D\tilde{D} + p_0 - \frac{1}{2} \gamma_2 \tilde{E}^2 = 0
\end{cases}
\end{align*}

These systems must be solved subject to the terminal conditions at $t=T$:
\begin{gather*}
A(T) = B(T) = C(T) = D(T) = E(T) = F(T) = 0 \\
\tilde{A}(T) = \tilde{B}(T) = \tilde{C}(T) = \tilde{D}(T) = \tilde{E}(T) = \tilde{F}(T) = 0
\end{gather*}

 \end{document}